\documentclass[aps,prl, showpacs,singlecolumn]{revtex4}
\usepackage{amssymb}
\usepackage{amsmath}
\usepackage{graphicx}
\usepackage{float}
\usepackage{dcolumn}
\usepackage{bm}

\begin{document}

\title{Controlling quantum state transfer in spin chain with the confined
field}
\author{Bing Chen$^{1}$, Z. Song$^{1,a}$ and C. P. Sun $^{1,2,a,b}$}
\affiliation{$^{1}$Department of Physics, Nankai University, Tianjin 300071, China}
\affiliation{$^{2}$ Institute of Theoretical Physics, Chinese Academy of Sciences,
Beijing, 100080, China}

\begin{abstract}
As a demonstration of the spectrum-parity matching condition (SPMC) for
quantum state transfer, we investigate the propagation of single-magnon
state in the Heisenberg chain in the confined external tangent magnetic
field analytically and numerically. It shows that the initial Gaussian wave
packet can be retrieved at the counterpart location near-perfectly over a
longer distance if the dispersion relation of the system meets the SPMC
approximately.
\end{abstract}

\maketitle

\section{Introduction}
Transferring a quantum state or object from one place to another
is an important task in quantum information science. Recent works
find that the quantum spin chain can be used to act as
\textquotedblleft quantum data bus\textquotedblright\ to perfectly
transfer quantum states which has drawn
the attention of quantum communication community \cite%
{Bose1,Liying,Song,Christandle1,Christandle2,Christandle3,Bose2,Bose3,Shitao1,Shitao2}%
. A great advantage of this approach is that no dynamical controls are
needed after we prepared the quantum state (except for the detection of the
state at the other end of spin chain) that is only the sending and the
receiving parties apply gates to the system, but the part of the chain
interconnecting them cannot be controlled during the communication process.

In the first proposal \cite{Bose1}, Bose considered a regular
one-dimensional spin chain with Heisenberg interactions which is able to
transfer quantum state over reasonable distance with the aid of distillation
process. This proposal is not the perfect scheme to transfer quantum state
due to the dispersion of the spin chain. This result in the fidelity for a
single-qubit state transfer becomes worse as the chain get longer.

Since then, a number of interesting proposals are proposed for quantum
communication through spin system to improve the fidelity of the quantum
information transfer. One of them is to choose proper the modulation of the
coupling strengths between two nearest neighbor sites as suggested in \cite%
{Christandle1,Christandle2,Christandle3}. In such system, although
an arbitrary local quantum state will spread as the time evolves,
after a period of time the dispersed information will
\textquotedblleft refocus\textquotedblright\ at the receiving
location of the chain. So perfect states transfer can be realized.
Another approach is making use of gapped systems. The advantage of
these schemes is the intermediate spins are only virtually excited.
In this case, the two separated qubits are coupled and realized the
entanglement of two point \cite{Liying}. It ensures that the
transfer of single-qubit state has a very high fidelity.

According to quantum mechanics, perfect state transfer can be
realized in many continuous and discrete systems. For simplicity, we
only consider a single-particle system with the usual spatial
refection symmetry (SRS). The SRS can be expressed by $[H,P]=0$,
where $P$ is the spatial refection
operator and $H$ is the Hamiltonian of the system. If the eigen values $%
\varepsilon _{n}$ match the parities $P_{n}$ in the following way
\begin{equation}
\varepsilon _{n}=N_{n}E_{0},P_{n}=\pm (-1)^{N_{n}},  \label{spmc}
\end{equation}%
any state $\varphi (\vec{r})$ can evolve into the reflected state
\begin{equation}
P\varphi (\vec{r})=\pm \varphi (-\vec{r})
\end{equation}
after time $\tau =(2n+1)\pi /E_{0}$, where $\vec{r}$ is the position of
sites in the lattice. Eq. (\ref{spmc}) is called the spectrum-parity
matching condition (SPMC), which can conduct us to find various artificial
solid state system transferring quantum states with high fidelity. It is
easy to find that the model of \cite{Christandle1} corresponds to the SPMC
for the simplest case $N_{n}=n$.

As analyzed above, many different schemes can realize a quantum state
transfer between two parties by means of a spin chain: the state of the
leftmost qubit is transferred to the rightmost qubit after a given time,
depended on the dynamics of the chain, to achieve perfect state transfer
over arbitrary distances. It also indicates that the revival period is
determined by the size of the system, but not the distance between two
parties. So it is necessary for us to find a new model to control the
revival period. On the other hand, the model we considered must satisfy the
SPMC. The previous work \cite{Shitao2} has considered $(2N+1)$-site spin-$%
1/2 $ ferromagnetic Heisenberg chain in the parabolic magnetic field, which
is the most simple model satisfying SPMC. Inspired by this paper, it is
natural for us to study other extension of the application of the theorem
with $N_{n}\neq n$. In this paper, we present a new example, based on the
SPMC and super-symmetry theory in quantum mechanics. The protocol we
considered is a ferromagnetic Heisenberg chain with uniform coupling
constant in an external magnetic field. It has been shown \cite{Shitao1,
Shitao2} that if the external field is parabolic, the lower spectrum is
equal-spacing, which is a good example to meet SPMC in the simplest case $%
N_{n}=n$. The super-symmetry theory in quantum mechanics \cite{SS} implies
that if the applied field is tangent-shaped, the lower spectrum should
satisfy the SPMC in the case $N_{n}=n^{2}$. In the following sections, we
will investigate the propagation of single-magnon state in the Heisenberg
chain in the confined external tangent magnetic field analytically and
numerically.

\section{The scheme and model}
In general, the transfer of a qubit state from the location $A$ to
$B$ can be regarded as the following process. The initial qubit
state $\left\vert \psi _{A}\right\rangle =u\left\vert
1\right\rangle _{A}+v\left\vert 0\right\rangle _{A}$ is prepared
at $A$, where $\left\vert 1\right\rangle _{A}$ and $\left\vert
0\right\rangle _{A}$ denote the two-level basis of the qubit at
$A$. For a spin system, we have $\left\vert 1\right\rangle
_{A}=\left\vert \uparrow \right\rangle _{A}$\ and $\left\vert
0\right\rangle _{A}=\left\vert \downarrow \right\rangle _{A}$.\ If
one can find an operation $U_{AB}$\ to realize
\begin{equation}
U_{AB}\left\vert \psi _{A}\right\rangle =u\left\vert 1\right\rangle
_{B}+e^{i\varphi _{AB}}v\left\vert 0\right\rangle _{B}
\end{equation}%
where $\varphi _{AB}$\ is the known phase for a given system, we say the
qubit state is transferred from $A$ to $B$ perfectly. Bose \cite{Bose1}
pointed out that the operation $U_{AB}$\ can be achieved in the qubit array
by the time evolution of the system. Especially, the perfect state transfer
can be performed if the system meets the SPMC. Unfortunately, such kind of
array needs a deliberated modulation \cite%
{Christandle1,Christandle2,Christandle3, Shitao2} for the couplings between
two qubits, which is difficult to be pre-engineered in experiment.
Nevertheless, if the transferred state is not a qubit state at a certain
site but the superposition of the single-qubit state localized in a small
range of coordinate space, e.g., a Gaussian wave packet (GWP) represented as
\begin{equation}
\left\vert \psi _{A}\right\rangle =C\sum_{i}\exp \left[ -\frac{1}{2}\alpha
^{2}(i-N_{A})^{2}\right] (u\left\vert 1\right\rangle _{i}+v\left\vert
0\right\rangle _{i})\prod\limits_{j\neq i}\left\vert 0\right\rangle _{j},
\label{gwp}
\end{equation}%
where $C$ is the normalization factor, it can be transferred to a
certain position $N_{B}$\ near perfectly \cite{Shitao2}. It
indicates that, for a
spin model with the saturated ferromagnetic state $\prod\nolimits_{j}\left%
\vert 0\right\rangle _{j}$ being the eigenstate, the transfer of a
single-magnon GWP is equivalent to that of a qubit.

In this section, we propose a model whose spectrum structure obeys the SPMC
in low energy for the near perfect transfer of a GWP. Let us consider the
Hamiltonian of $(2N+1)$-site spin-$1/2$ ferromagnetic Heisenberg chain
\begin{eqnarray}
H &=&-J\sum_{i=-N}^{N-1}\overrightarrow{S}_{i}\cdot \overrightarrow{S}%
_{i+1}+\sum_{i=-N}^{N}B(i)S_{i}^{z}  \notag \\
&=&-J\sum_{i=-N}^{N-1}\left[ \frac{1}{2}%
(S_{i}^{+}S_{i+1}^{-}+S_{i}^{-}S_{i+1}^{+})+S_{i}^{z}S_{i+1}^{z}\right]
+\sum_{i=-N}^{N}B(i)S_{i}^{z},  \label{H}
\end{eqnarray}%
where $J>0$ is the exchange coupling constant. The external magnetic field
has the form%
\begin{equation}
B(i)=B_{0}\left[ \tan ^{2}\left( \pi \eta i\right) +\frac{1}{2}\right] ,
\label{B(i)}
\end{equation}%
where $B_{0}=2\lambda J\pi ^{2}\eta ^{2}$, and $L_{eff}=1/\eta $ denotes the
effective length of the system, indicating the confined range of the single
magnon. Here $S_{i}^{x}$, $S_{i}^{y}$ and $S_{i}^{z}$ are Pauli matrices for
$i$-th site. Obviously, for the Hamiltonian Eq. (\ref{H}) the $z$-component
of the total spin%
\begin{equation}
S^{z}=\sum_{i=-N}^{N}S_{i}^{z}
\end{equation}%
is the conservative quantity of the system, i.e.,
\begin{equation}
\left[ H,S^{z}\right] =0.
\end{equation}%
Then in the invariant subspace with the fixed $S^{z}=N-1/2$, the Ising term
in the Hamiltonian $J\sum_{i=-N}^{N-1}S_{i}^{z}S_{i+1}^{z}$ just contributes
a constant to the matrix of Hamiltonian, which can be neglected. With the
aid of Jordan-Wigner transformation \cite{J-W} the single-magnon effective
Hamiltonian can be expressed as a spin-less fermion lattice model with a
tan-potential
\begin{equation}
H_{eff}=-\frac{J}{2}\sum_{i=-N}^{N-1}\left( a_{i}^{\dag }a_{i+1}+h.c.\right)
+B_{0}\sum_{i=-N}^{N}\tan ^{2}\left( \pi \eta i\right)   \label{Heff}
\end{equation}%
where $a_{i}^{\dag }$ is the fermion operator at $i$-th site, $%
B_{0}=2\lambda J\pi ^{2}\eta ^{2}$ and we have neglected a constant for
simplicity. The spectrum of Eq. (\ref{Heff})\ can not be solved exactly
except the case of very small $N$.\ However, the corresponding continuous
model\ has exact solution and can give us some implications. Actually,
consider the Hamiltonian Eq. (\ref{Heff}) in $B_{0}=0$\ case, the
tight-binding spectrum is
\begin{equation}
\epsilon _{k}=-2J\cos k,
\end{equation}%
where $k=n\pi /\left( 2N+2\right) $, $n=1,2,...,2N+1$. Notice that in lower
energy region $k\sim 0$\ the spectrum is $\epsilon _{k}\approx $\ $%
-2J(1-k^{2})$\ $\sim 2Jk^{2}$, which is very close to that of free particle
in a continuous system, i.e., $E=p^{2}/2m$. It implies that for lower energy
scale, the spectrum can be obtained from the corresponding continuous model
approximately. According to quantum mechanics \cite{SS}, the energy spectrum
of the continuous model
\begin{equation}
H_{con}=2Jp^{2}+B_{0}\tan ^{2}\left( \pi \eta x\right)
\end{equation}%
is%
\begin{equation}
E_{n}=J\pi ^{2}\eta ^{2}\left( n^{2}+4\mu n+C\right)   \label{En}
\end{equation}%
\ where
\begin{eqnarray}
\mu  &=&\frac{1}{4}\left( \sqrt{8\lambda +1}-1\right) ,  \notag \\
C &=&J\pi ^{2}\eta ^{2}\left( \sqrt{8\lambda +1}-1\right) .
\end{eqnarray}%
The reason to choose the tan-potential is due to its rich spectrum structure
as $\lambda $\ varies. For $\lambda =1$, the corresponding energy spectrum is%
\begin{equation}
E_{n}=\frac{1}{2}n(n+2)B_{0},\text{ }n=1,2,\cdots .
\end{equation}%
\ which can be written as the quadratic term of $(n+1)$.\textrm{\ }On the
other hand, when $\lambda \gg 1$\ is taken into account, the linear term
becomes dominant, which corresponds to that of parabolic potential \cite%
{Shitao2}.

In this paper, we mainly focus on the case of $\lambda =1$, since the
quadratic dispersion also meet the SPMC. For a given $L_{eff}=500$ as an
example, the magnitude of magnetic field is $B_{0}/\lambda =7.9\times
10^{-5}J$. The system is illustrated in Fig. 1.
\begin{figure}[tbp]
\includegraphics[bb=80 236 535 532, width=8 cm, clip]{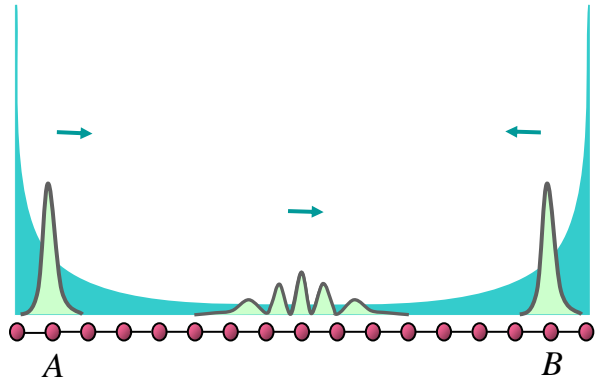}
\caption{Schematic figure of the system with tan-potential
$B(x)=B_{0}\tan ^{2}(\protect\pi \protect\eta x)$ indicated by the
curve line. The initial GWP localizes at the point $A$. After
$t=2\protect\pi /B_{0}$, the state will be retrieved at $B$ as
time evolves.} \label{fig1}
\end{figure}
Obviously, such spectrum meets to the SPMC for the case $N_{n}=n(n+2)$.
According to SPMC, any initial state $\psi \left( x\right) $ with dominant
components in low energy scale can be evolved into $\pm \psi \left(
-x\right) $ at time $\tau =2(2n+1)\pi /B_{0}$ approximately. On the other
hand, since the saturated ferromagnetic state $\prod\nolimits_{j}\left\vert
0\right\rangle _{j}$ is the ground state of the model, which is unchanged
during the time evolution. Then if a GWP of a single-magnon state can be
transferred near perfectly, this model can be employed to implement quantum
information transmission in solid-based device.

\begin{figure}[tbp]
\includegraphics[bb=19 343 525 744, width=7 cm, clip]{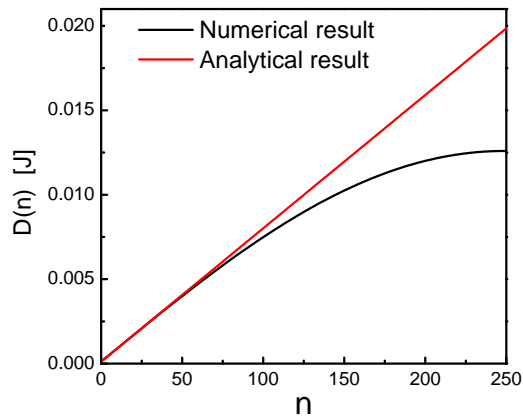}
\caption{The energy spacing $D(n)=E_{n+1}-E_{n}$ (in the unit of $J$)
obtained by numerical simulation (black line) for $L_{eff}=500$ and $B_{0}=2J%
\protect\pi ^{2}\protect\eta ^{2}$. The red line is the plot of Eq. (\protect
\ref{En}). It shows that at low energy region the spectrum of single magnon
is quadratic in high accuracy. }
\label{fig2}
\end{figure}

In order to demonstrate the analysis of the spectrum of the system,
numerical simulation is performed for the system with $L_{eff}=500$. In Fig. %
\ref{fig2}, we show the numerical and analytical result of energy spacing $%
D\left( n\right) =E_{n+1}-E_{n}$\ (in the unit of $J$) as the function of
energy level $n$. It shows that there is indeed a liner region for lower
energy levels (about $n<80$), which satisfy the SPMC approximately. So if
the GWP can be expanded by the lower eigenstates dominantly, it can be
transferred near perfectly.

\section{Quantum transfer of GWP}
In this section, we show some numerical examples of quantum state
transmission in the spin chain with the uniform coupling strength
$-J<0$, controlled in the magnetic field in the form of Eq.
(\ref{B(i)}). We
consider a GWP with the center at $N_{A}$ as the initial state at\textbf{\ }$%
t=0$. The packet we choose must be approximately expanded by the energy
levels, which satisfy the SPMC.
\begin{equation}
\left\vert \psi (N_{A},0)\right\rangle =C\sum_{i=-N}^{N}\exp \left[ -\frac{1%
}{2}\alpha ^{2}(i-N_{A})^{2}\right] \left\vert i\right\rangle
\end{equation}%
where $\left\vert i\right\rangle =\left\vert 1\right\rangle
_{i}\prod\nolimits_{j\neq i}\left\vert 0\right\rangle _{j}$ denotes the
state with a single spin flip at $i$-th site. The factor $\alpha $
determines the width of the wave-packet, $\Delta =2\sqrt{\ln 2}/\alpha $. At
time $t$ the state $\left\vert \psi (N_{A},0)\right\rangle $ evolves to
\begin{equation}
\left\vert \phi (t)\right\rangle =e^{-iH_{eff}t}\left\vert \psi
(N_{A},0)\right\rangle .
\end{equation}%
For small $\alpha $,\ $\left\vert \psi (N_{A},0)\right\rangle $\ can be
expanded by the eigenstates $\left\vert \varphi _{n}\right\rangle $\ of $%
H_{eff}$,\ which belong to the quadratic region, i.e.,
\begin{equation}
H_{eff}\left\vert \varphi _{n}\right\rangle \simeq E_{n}\left\vert \varphi
_{n}\right\rangle .
\end{equation}%
Thus at $\tau =2\pi /B_{0}$ the final state is
\begin{eqnarray}
\left\vert \phi (t=\tau )\right\rangle &=&e^{-iH\tau }\left\vert \psi
(N_{A},0)\right\rangle  \notag \\
&\simeq &\sum_{n}e^{-iE_{n}\tau }\left\vert \varphi _{n}\right\rangle
\left\langle \varphi _{n}\right\vert \left. \psi (N_{A},0)\right\rangle
\notag \\
&=&\sum_{n}e^{-i\pi n(n+2)}\left\vert \varphi _{n}\right\rangle \left\langle
\varphi _{n}\right\vert \left. \psi (N_{A},0)\right\rangle  \notag \\
&=&P\left\vert \psi (N_{A},0)\right\rangle  \notag \\
&=&\left\vert \psi (-N_{A},0)\right\rangle ,
\end{eqnarray}%
which is the reflection counterpart of the initial one.

As for quantum information, what we concern is the fidelity of qubit-state
transmission. Since the saturated ferromagnetic state $\prod\nolimits_{j}%
\left\vert 0\right\rangle _{j}$\ is the ground state of the Hamiltonian, and
belongs to different subspace from the single-magnon state, the fidelity of
quantum information is determined by the fidelity of the GWP transmission
\begin{equation}
F(t)=\left\vert \left\langle \psi (-N_{A},0)\right\vert
e^{-iH_{eff}t}\left\vert \psi (N_{A},0)\right\rangle \right\vert .
\end{equation}
On the other hand, the above analytical result is obtained in an approximate
manner. In order to investigate the accuracy of the approximation for
various systems, numerical simulation is employed. For this purpose, we
numerically solve the Schr\"{o}dinger equation for the dynamical evolution
and compute the fidelity of the wave-packet transmission from rightmost $%
N_{A}$\ of the spin chain to the leftmost $-N_{A}$.

\begin{figure}[tbp]
\includegraphics[bb=17 62 531 753, width=7 cm, clip]{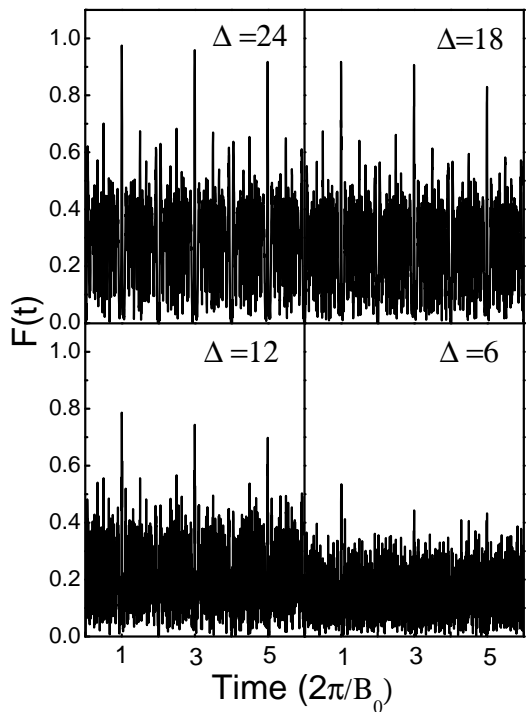} .
\caption{Fidelity of the GWPs with $\Delta =24$, $18$, $12$, and $6$
transferred over the distance of $L=2N_{A}$ $=400$\ in the system with $%
L_{eff}=500$.}
\label{fig3}
\end{figure}

\begin{figure}[tbp]
\includegraphics[bb=31 302 510 750, width=7 cm, clip]{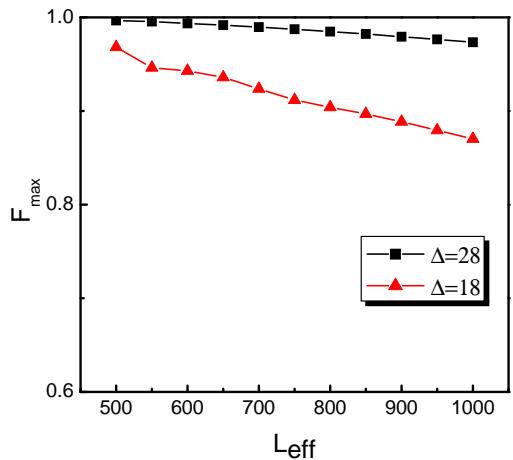}
\caption{The plot of maximal fidelity of the GWPs with $\Delta =28$, $18$
transferred over the distance of $L=2L_{eff}-100$ in the system with $%
L_{eff}\in \lbrack 500,1000]$. It shows that, as the effective length gets
longer, the fidelity slightly decreases if the width of GWP is taken
appropriately.}
\label{fig4}
\end{figure}
As mentioned before, the SPMC can be satisfied in the range of
lower spectrum. The width of GWP $\Delta $\ determines the range
of the levels span its wave function, as well as the fidelity of
the transmission. In other word, the smaller $\Delta $\ is, the
lower the fidelity will be. We consider the transmissions of the
GWP with $\Delta =24$, $18$, $12$, and $6$ over the distance
$L=2N_{A}$ $=400$\ in the system with $L_{eff}=500$. The
fidelities as function of time $t$\ are plotted in Fig.
\ref{fig3}. We find that the fidelities get their maxima around
$(2n+1)\tau $ $(n=0,1,2,...)$, which are in agreement with the
analytical prediction. It shows that the maximal fidelity becomes
high as the width of GWP increases. Then for the scheme of quantum
information transfer, the higher fidelity can be achieved if the
GWP with $\Delta \geq 24$ is employed. Interestingly, one can see
that there exists a regular "noise" in the plot of the fidelity.
This
phenomenon is related to the fractional revival of the wave-packet \cite%
{Fractional revival} and will be discussed in the further work.

On the other hand, we also investigate the influence of the transfer
distance to the fidelity numerically. Numerical simulation is
performed to obtain the maximal fidelity $F(t)$\ of transmission of
GWPs with $\Delta =$\ $18$ and $28$ over the distance of $L=2N_{A}$
$=L_{eff}-100$ in the systems with $L_{eff}\in \lbrack 500,1000]$.
In Fig. \ref{fig4}, the maximal fidelity as a function of the
confined length $L_{eff}$\ is plotted. It shows that, as the
effective length gets longer, the fidelity slightly decreases if the
width of GWP is taken appropriately. It indicates that this scheme
can be a good candidate to realize quantum information transmission
in solid-based device.

\begin{figure}[tbp]
\includegraphics[bb=34 344 536 740, width=7 cm, clip]{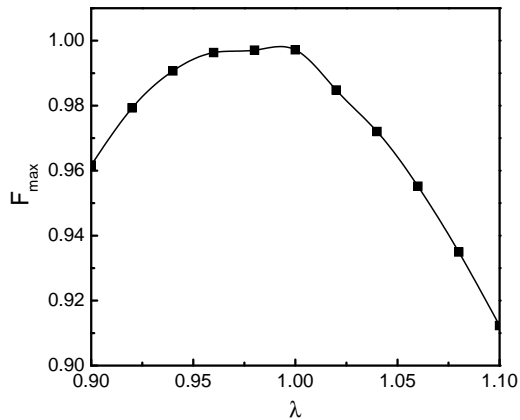}
\caption{Maximum fidelity of the GWPs with $\Delta =24$ transferred
over the distance of $L=400$ in the system with $L_{eff}=500$, as a
function of $\protect\lambda $. It indicates that $\protect\lambda
\lesssim 1 $ is approximately optimal.} \label{fig5}
\end{figure}

In the above we focus on the case of $\lambda =1$\ due to the
indication from the continuous system. A natural question is asked
whether $\lambda =1$ is the optimal for the GWP transmission in the
vicinity of $1$. Numerical study has been done to calculate the
fidelity as a function of $\lambda $\ for the propagation of a GWP
with $\Delta =24$ over the distance of $2N_{A}=400$\ \ in the system
with $L_{eff}=500$. The result is plotted in Fig. \ref{fig5}, which
shows that $\lambda \lesssim 1$ is approximately optimal.

\section{Harmonic approximation in strong field limit}
All above discussions are based on the tan-potential with $\lambda
=1$ and numerical result has demonstrated that it is the optimal
potential for GWP transmission. In this section, we concentrate on
another special case of strong field limit. From the above
analysis, we know that the low spectrum for arbitrary value of
$\lambda $ is similar to Eq. (\ref{En}), which is the combination
of linear and quadratic dispersions. Obviously, it is easy to find
that the linear dispersion is dominant when we consider the strong
field limit, $\lambda \gg 1$. In other word, under the condition
of the strong field limit, the effective spectrum for the GWP is
quasi-harmonic. In this case the SPMC is also satisfied and can be
employed to transfer quantum information \cite{Shitao2}.
\begin{figure}[tbp]
\includegraphics[bb=54 340 525 741, width=7 cm, clip]{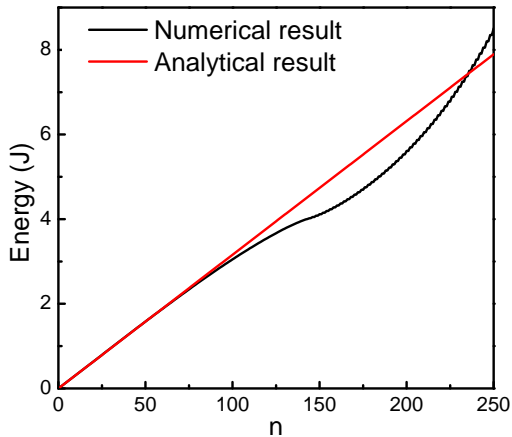}
\caption{The energy spectrum (in the unit of $J$) obtained by numerical
simulation (black line) for $L_{eff}=500$ and strong field with $B_{0}=6.33J$%
. The red line is the plot of $E_{n}^{\prime }=0.032Jn$. It shows that at
low energy region the spectrum of single magnon is linear in high accuracy.}
\label{fig6}
\end{figure}
\begin{figure}[tbp]
\includegraphics[bb=14 47 577 810, width=7.0cm, clip]{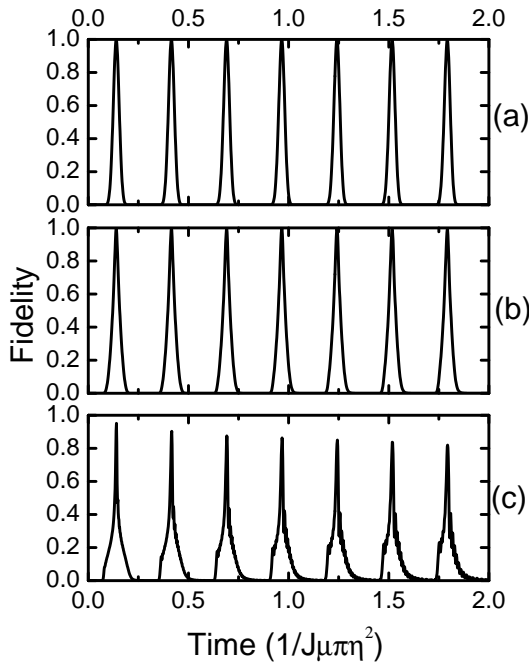}
\caption{Fidelity of the GWPs with (a) $\Delta =6$; (b) $\Delta =4$; (c) $%
\Delta =2$, transferred over the distance of $L=120$ in the system with $%
L_{eff}=500$ and $B_{0}=6.33J$, as a function of time $t$.}
\label{fig7}
\end{figure}
\begin{figure}[tbp]
\includegraphics[bb=16 48 575 809, width=7.0cm, clip]{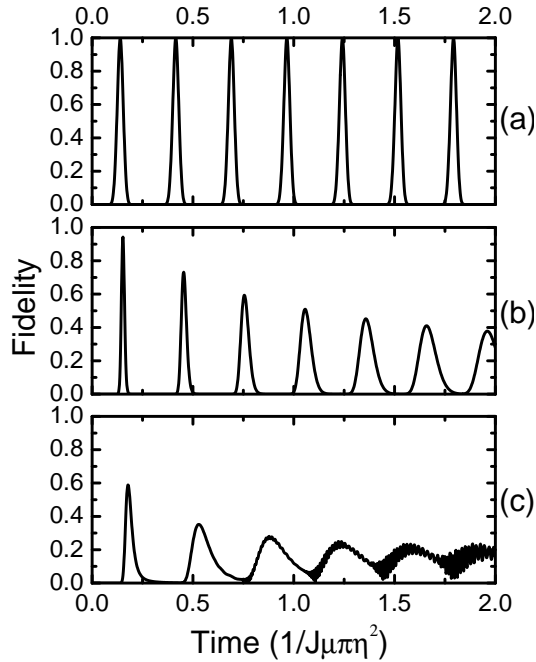}
\caption{Fidelity of the GWPs with $\Delta =6$, transferred over the
distance of (a) $L=120$; (b) $L=200$; (c) $L=220$ in the system with $%
L_{eff}=500$ and $B_{0}=6.33J$, as a function of time $t$.}
\label{fig8}
\end{figure}
In order to demonstrate the above analysis, we perform numerical
simulation
for the system with strong field $B_{0}=$ $6.33J$ (or $\lambda =$ $%
8.02\times 10^{5}$) and $L_{eff}=500$. The energy spectrum of such system
with obtained by exact diagonalization and the analytical result from the
continuous system in large $\lambda $\ limit, i.e.
\begin{equation}
E_{n}^{\prime }=2\sqrt{2\lambda }J\pi ^{2}\eta ^{2}n=0.032Jn,
\end{equation}%
are plotted in Fig. \ref{fig6}. It shows that the lower spectrum (about $%
n\simeq 80$) is equal-spacing in high accuracy. In Fig.
\ref{fig7}, we plot the fidelity as a function of time for the
propagation of GWP with different half-width $\Delta =2,$ $4,$ and
$6$ over the distance of $L=120$. It shows that, the
quasi-harmonic potential is more appropriate to transfer
wave-packet with small width and the fidelity is higher. Meanwhile
the recurrent period is shorter. On the other hand, we also
investigate the influence of the transfer distance $L$ to the
fidelity. From Fig. \ref{fig8} we can see that the fidelity
quickly decays as the transfer distance gets longer.

\section{Conclusion}
We have studied analytically and numerically, the propagation of
single-magnon state in the Heisenberg chain in the confined
external tangent magnetic field with different strength. We find
that it can be a good model to demonstrate the SPMC for quantum
state transfer in spin systems. It is shown that the initial
Gaussian wave packet can be retrieved at the counterpart location
near-perfectly over a longer distance if the dispersion relation
of the system meets the SPMC approximately. It also shows that the
qubit array with appropriate external field is a suitable scheme
for the task of quantum information transmission in solid-state
system.

This work is supported by the CNSF (grant No. 10474104), the National
Fundamental Research Program of China (No. 2001CB309310).

\end{document}